\documentclass[preprint,showpacs,preprintnumbers,amsmath,amssymb,nofootinbib]{revtex4}

\usepackage{graphicx}
\usepackage{dcolumn}
\usepackage{bm}

\begin{document}

\title{SyZyGy: A Straight Interferometric \\
Spacecraft System for Gravity Wave Observations}

\author{F. B. Estabrook}
\email{Frank.B.Estabrook@jpl.nasa.gov}
\affiliation{Jet Propulsion Laboratory, California Institute of Technology,
 Pasadena, CA 91109}

\author{J. W. Armstrong}
\email{John.W.Armstrong@jpl.nasa.gov}
\affiliation{Jet Propulsion Laboratory, California Institute of Technology,
 Pasadena, CA 91109}

\author{Massimo Tinto}
\email{Massimo.Tinto@jpl.nasa.gov}
\altaffiliation [Also at: ]{Space Radiation Laboratory, California
  Institute of Technology, Pasadena, CA 91125}
\affiliation{Jet Propulsion Laboratory, California
Institute of Technology, Pasadena, CA 91109}

\author{William Folkner}
\email{William.Folkner@jpl.nasa.gov}
\affiliation{Jet Propulsion Laboratory, California
Institute of Technology, Pasadena, CA 91109}

\date{\today}
\vskip24pt

\begin{abstract}

We consider a spaceborne gravitational wave (GW) detector formed
by three spacecraft in a linear array, coherently exchanging laser
beams and using the data combinations of time-delay interferometry
(TDI).  We
previously showed how the measured time series of Doppler shifts in
the six one-way laser links between spacecraft pairs in a general
unequal-arm triangular configuration can be combined, using TDI, to
exactly cancel the otherwise overwhelming phase noise of the lasers
while retaining sensitivity to GWs.  Here we apply TDI, unfolding the
general triangular configuration, to the special case of a linear array
of three spacecraft.  We show that such an array (``SyZyGy'') has,
compared with an equilateral triangle GW detector of the same scale,
degraded (but non-zero) sensitivity at low-frequencies (f $<<$ c/(array
size)) but similar peak and high-frequency sensitivities to GWs.  We
develop the GW and noise responses, noting that in this geometrical
case only one TDI combination is GW-sensitive showing a
relatively simple 6-pulse response to an incident GW burst.
Sensitivity curves are presented for SyZyGys having various
arm-lengths.  A number of technical simplifications result from the
linear configuration.  These include:  only one faceted (e.g. cubical) proof mass per
spacecraft, intra-spacecraft laser metrology needed only at the central
spacecraft, placement in an appropriate orbit can reduce Doppler
drifts so that no laser beam modulation is required for ultra-stable
oscillator noise calibration, and little or no time-dependent
articulation of the telescopes to maintain pointing.  Because SyZyGy's
sensitivity falls off more sharply at low-frequency than that of an
equilateral triangular array, it may be most useful for GW observations
in the band between those of ground-based interferometers
($\sim 10 - 2000$ Hz) and LISA ($\sim 10^{-4} - 10^{-1}$ Hz).  SyZyGy
with $\sim$ 1 light-second scale could, for the same instrumental
assumptions as LISA,  make observations
in this intermediate frequency GW band with $5 \sigma$
sensitivity to sinusoidal waves $\simeq 2.5 \times 10^{-23}$ in a year's
integration.

\end{abstract}


\maketitle

\section{Introduction}

When coherent optical or microwave wave trains are interchanged between
three or more freely orbiting spacecraft, and phase or frequency
differences of all incoming and outgoing beams recorded, there is a
closure that allows subsequent elimination of all source fluctuations
(``laser noises'') and so achieves sensitivity to propagating
gravitational waves at much lower levels set by secondary noises. In
doing this, the precise propagation times between sources and readout
locations enter the data reduction algorithms, but equality of arm
lengths is not necessary.  We have called this scheme ``time delay
interferometry'' or TDI, \cite{1} and subsequently analyzed several
configurations and data types needed for further elimination of the
most serious secondary phase noise sources, {\it{viz.}} those due to
spacecraft non-geodesic motion (``optical bench noise'') and especially
to fluctuations of the ultra-stable local oscillators, or frequency
standards, used in fringe tracking (``USO noise'') \cite{2,3}.  Fringe
tracking is necessary when the lasers in the configuration have widely
differing frequencies, and/or when the configuration does not move
rigidly so that emitted and received signals are Doppler shifted during
slow spacecraft separation changes.

The LISA space gravitational wave antenna will use TDI, and is to be
roughly equilateral, which maximizes the overall response to low
frequency gravitational waves over a wide band.  
Reduction of secondary noise,  and simplicity of system architecture,
nevertheless suggest consideration also of other schemes;  here we
consider the most extreme possible application of TDI, {\it{viz.}} to a
linear orbiting array, or, in astronomical terms, three spacecraft
flying in syzygy\footnote{An amusing coincidence is that Dhurandhar,
Nayak and Vinet \cite{4} have pointed out that underlying the data
combinations of TDI is the mathematics of polynomial rings in three
variables, and that the structure we found is called by algebraists a
``module of syzygies''!  SyZyGy for this reason again would seem to be
an apt acronym for the concept of this paper.}.

We have previously considered a linear array, or parallel beam
interferometric detector of gravitational waves, in work showing that
with multiple bounces a split antiparallel laser beam will amplify a
gravitational wave signal while laser fluctuations are cancelled
conventionally by precise control of equal arm lengths \cite{4}.  We
now see the multiple readout scheme of TDI as much preferable, and
equal spacing of the three spacecraft is no longer required.  The
equations for combining the heterodyne data streams in TDI can be taken
directly from our previous papers, which had general triangular
configuration; in this paper they will be used in the linear limit.

The (nearly) linear formation of SyZyGy has some advantages and
disadvantages over the nominal equilateral triangular formation planned
for the LISA mission. The primary disadvantages of the linear formation
are less sensitivity at the low frequencies (for a given spacecraft
separation distance) and sensitivity to only one linear combination of
GW polarizations.  The advantages arise from the potential for a more
stable formation geometry. The nominal LISA triangular formation
involves spacecraft in orbits with semi-major axis a, inclination I, and
eccentricity e related by $e = I/3^{1/2}$ and $I = L/(2a)$ \cite{4aa,
4bb}.  This choice of orbital elements produces a rotating triangular
configuration with arm lengths that are constant to $\sim e^2 L$. The
time variation of the arm lengths causes a variable Doppler shift on
the laser beam signals, a variable direction from each spacecraft to
the other two spacecraft, and a variable angle between the spacecraft
velocity and the direction to the other spacecraft.  The (almost) linear array
might be formed with three spacecraft moving in a single circular orbit about the
sun.  In the ideal case (ignoring perturbations from other bodies) the
distance between spacecraft and the directions from each spacecraft to
the others will remain constant.  The constant distance between
spacecraft removes the Doppler shift on the laser signals. The constant
directions means that no mechanism is needed to account for angle
variation.  The constant angle also means that a single faceted (e.g.
cubical) test mass can be used in each spacecraft, instead of the two
in the nominal LISA configuration. (In the case of spherical test
masses, the constant angle between the velocity direction and direction
to other spacecraft removes some surface irregularity errors that would
affect a triangular formation with spherical test masses.) The constant
angle between the velocity direction and direction to other spacecraft
means that the point-ahead and point-back angles are constant. (These angles
arise from the need to point lasers to hit the location the other spacecraft
will be at after a one-way light time.) 
The
linear combination needs less spacecraft propulsion to reach the
desired orbits (for a given location relative to Earth and arm length
L) since no plane change is needed.  No laser metrology is needed for
the proof masses relative to the optical benches at the outlying
spacecraft (e.g., data we have denoted $z_{12}$, $z_{32}$, at
spacecraft 2).  (At the center spacecraft (spacecraft 1) the
intra-spacecraft metrology data denoted $z_{21}$ (or $-z_{31}$) in the
TDI equations must still be taken to eliminate phase fluctuations due
to vibrations of the optical benches.)

%
%

The procedure for elimination of USO noise introduces considerable
complication in the LISA system, inasmuch as additional calibrating
data must be taken and exchanged if present generation flight-qualified
USO's are used.  The USO noise enters the TDI combinations multiplied
by the internally generated fringe tracking frequencies $\omega_{ij}$
that are required when there are inter-spacecraft Doppler drifts and/or
offsets of the laser center frequencies. Control of relative spacecraft
motions will automatically be achieved by spacing SyZyGy along a single
nearly-circular solar orbit, and laser frequency offsets can be limited
either by atomic line stabilization or by use of optical transponders
\cite{4a}.  Thus no wide-band fringe tracking data, USO calibration,
and associated modulation of the laser beams are needed by SyZyGy, as
will be shown in the next section.

In the linear case the gravitational wave antenna patterns of all the
TDI data combination become simple and identical (so-called 6-pulse
response).  We give this pattern explicitly in Section 2.  The pattern
will of course sweep over the sky as the spacecraft move along their
common orbit. We apply the TDI data combinations in the linear
geometry, and show how the only data combinations that are sensitive to
GWs have the same signal response as the combination X, the unequal arm
Michelson interferometer.  We introduce two new independent
combinations that have no gravitational wave response, denoted $\rho$
and $\sigma$;  they can be used to monitor system noises.  All
previously identified laser-noise-free data combinations are given in
terms of X, $\rho$ and $\sigma$.

In Section 3 we derive the SyZyGy response functions for the remaining
secondary noises:  optical bench vibration, proof mass residual
accelerations, optical path noises at the photodetector readouts, and
USO noise entering in to the heterodyne measurements.  In Section 4 we
give numerical calculations and plots for these.  Averaging over the sky
and polarization states,
we finally plot the resulting gravitational wave
sensitivities to incident sine waves, using one year observation time
and several SyZyGy scale sizes.  Compared with an equilateral LISA, a
SyZyGy of the same scale has similar best and high-frequency
gravitational wave sensitivities.  However the sensitivity is poorer at
low-frequencies, suggesting applicability in the frequency band between
those of ground-based interferometers and LISA.

\section{Analysis}

We use the notation and results of \cite{2,3}.  A straight
interferometric array, shown in Figure 1, is obtained by opening up the general triangular
configuration of Figure 2 of reference \cite{2}, so that spacecraft 1
is in the center and $L_1 = L_2 + L_3$.  We denote the linear
orientation by the unit vector $\hat n$ = $\hat n_1 = - \hat n_2 = -
\hat n_3$.  On each of the spacecraft the usual pair of optical benches
can be fused into one, so we put $\vec V^*_i = \vec V_i$.  At the
outlying spacecraft 2 and 3 both incoming beams will be referenced to
the same face of a single proof mass.  At the center spacecraft, number
1, TDI will require use of two opposed faces of a single proof mass.
For all we will have $\vec v^*_i = \vec v_i$ and the optical bench and proof
mass motion components that enter are $\vec V_i \cdot \hat n$ and $\vec
v_i \cdot \hat n$, respectively.  The intra-spacecraft data $z_{ij}$ at
the three proof masses consequently satisfy $z_{32} = z_{12}, z_{13} =
z_{23}$, and $z_{21} = -z_{31}$.  The first four of these then drop out
of all the laser-noise-free data combinations and so in fact need not
be recorded.  The last two remain, and so one of them needs to be
taken, which amounts to implementing precision laser metrology solely
on the central spacecraft.

When all these specializations are substituted into the data
combinations derived for general configurations, it is found that they
degenerate into essentially one, six pulse, response to incident
gravitational waves (we use X) plus two others that respond only
to system noises (we call them $\rho$ and $\sigma$).  All three will
further be USO-noise-free to acceptable limits if the laser frequency
offsets are limited by Doppler
effects due to relatively stable spacecraft motion in the
SyZyGy linear orbital configuration.

Specifically, the new laser-noise-free combinations we use for a linear
array are

$X = y_{32;3'22'} - y_{23;233'} + y_{31;22'} - y_{21;33'} + y_{23;2} - y_{32;3'} +y_{21} - y_{31} - z_{21} - z_{21;11'} + z_{21;22'} + z_{21;33'}$

$\rho = y_{13} - y_{23} - y_{31;2'} + z_{31;2'}$

$\sigma = y_{12} - y_{32} - y_{21;3} + z_{21;3}$

\noindent
where a semicolon subscripted index $i$ is understood to mean not only
time delay by $L_i$ but also multiplication by the Doppler
factor $(1 - \dot L_i/c)$, which in this case can be dropped\footnote{In more general
work \cite{S03,CH03} the TDI combinations have been modified for the case of a moving
array where time delays such as $L_1$ from spacecraft 2 to spacecraft 3 differ
from $L_1^{'}$ from spacecraft 3 to spacecraft 2, etc.  One then has for SyZyGy
$L_1 = L_2^{'} + L_3^{'}$ and $L_1^{'} = L_2 + L_3$.  The primes on the time-delay subscripts
in these equations for X, $\rho$, and $\sigma$ have been inserted for that case, but we
do not insist on that generality in any of the subsequent
discussion.}.  Note how
interchanging indices 2 and 3 takes $\rho$ and $\sigma$ into one
another.

The previously used TDI basic combinations in this linear
array limit (and without motion) are given in terms of X, $\rho,$ and $\sigma$:

$\alpha = X  + \rho_{;2} - \sigma_{;3}$

$\beta = -\sigma + \rho_{;1}$

$\gamma = \rho - \sigma_{;1}$

$\zeta = \rho_{;3} - \sigma_{;2}$

\noindent
Proof mass and shot noise response functions for X and $\rho$
(or $\sigma$) are given in Section 4.

\section{Noise Transfer Functions}

The proof-mass-plus-optical-bench assembly and laser beam paths for the
central spacecraft is show schematically in Figure 2.  The
photodetectors that generate the time series $y_{21}$, $y_{31}$, and
$z_{21}$ are as indicated.  The outgoing light from spacecraft 1 to
spacecraft 2 is routed from laser 1 on the optical bench using mirrors
and beam splitters; this beam does not interact with spacecraft 1's
proof mass.  Conversely, the incoming light beam from spacecraft 2 is
first bounced off the nearest face of proof mass 1 before being
reflected onto the photodetector, where it is mixed with light from
laser 1.  This time series is $y_{31}$ in Figure 2.  Analogously, the
time series $y_{21}$ is producted from light transmitted to and
received from spacecraft 3.  The relative motion of the optical bench
and proof mass is monitored by measuring the Doppler shift of the light
coming directly from the laser with the light bounced off one face of
the proof mass (this forms the metrology time series $z_{21} =
-z_{31}$, as indicated in Figure 2.

Figure 3 shows the light paths on spacecraft 3.  The outgoing light
beam from spacecraft 3 to spacecraft 1 and 2 is routed from laser 3 on
spacecraft 3's  optical bench using mirrors and beamsplitters; again,
this beam does not interact with spacecraft 3's proof mass.  The two
incoming light beams from spacecraft 1 and 2 are bounced off the proof
mass before bing reflected onto photodetectors where they are mixed
with light from laser 3.  The time series produced are $y_{23}$ and
$y_{13}$, as shown.  The configuration at spacecraft 2 is analogous,
producing $y_{12}$ and $y_{32}$.  $y_{13}$ and $y_{12}$ enter $\alpha$,
$\rho$, and $\sigma$, but not X.

The equations for the noises entering the $y_{ij}$ and $z_{21} =
-z_{31}$ can be developed from Figures 2 and 3.  Consider first the
situation on spacecraft 1.  The photodetector producing $y_{31}$
(moving with velocity $V_1$) reads that time series by mixing the beam
originating from the optical bench on spacecraft 2 sent in direction
$\hat n_3 = -\hat n$ (laser noise $C_2$ and optical bench motion $\vec
V_2$, delayed by propagation along $L_3$), after one bounce off the
proof mass ($\vec v_1$), with the local laser light ($C_1$).  The time
series $y_{21}$ is produced analogously.  The $z_{21}$ measurement is
from light originating at the laser ($C_1, \vec V_1$), bounced off
the left face of the proof mass ($\vec v_1$), and mixed with the direct
laser light ($C_1$).  For spacecraft 2 and 3, $y_{32}$, and $y_{23}$
are produced similarly.  Metrology time series on the end spacecraft
are not required.

We will see below that there is no gravitational wave contribution to
the data $\rho$ and $\sigma$.  It may however be desirable to measure
these to monitor system responses and secondary noises.  In that
case we would also need the data $y_{12}$ and $y_{13}$.

The $y_{ij}$ and the required metrology time series $z_{21}$ (=
$-z_{31}$), including gravitational wave signals and shot noises, can
be developed from the diagrams and by consulting the Appendix of Tinto
{\it{et al.}} \cite{3}:

\begin{eqnarray}
\nu_0 \ y_{21} & = & \ \ \nu_3 (1 + C_{3,2} - \hat n_2 \cdot \vec V_{3,2}) \Delta_2 (1 + 2 \hat n_2 \cdot \vec v_1 - \hat n_2 \cdot \vec V_1)
\nonumber \\
& & -\nu_1 (1 + C_1) - \omega_{21} (1 + Q_1) + \nu_0 \ y_{21}^{gw} + \nu_0 \  y_{21}^{opt. \ path}
\label{eq:1}
\end{eqnarray}

\begin{eqnarray}
\nu_0 \ y_{31} & = & \ \ \nu_2 (1 + C_{2,3'} + \hat n_3 \cdot \vec V_{2,3'}) \Delta_3 (1 - 2 \hat n_3 \cdot \vec v
_1 + \hat n_3 \cdot \vec V_1)
\nonumber \\
& & -\nu_1 (1 + C_1) - \omega_{31} (1 + Q_1) + \nu_0 \ y_{31}^{gw} + \nu_0 \  y_{31}^{opt. \ path}
\label{eq:2}
\end{eqnarray}

\begin{eqnarray}
\nu_0 z_{21} & = & 2 \nu_1 \hat n_3 \cdot (\vec v_1 - \vec V_1)  = - \nu_0 z_{31}
\label{eq:3}
\end{eqnarray}

\noindent
where $\Delta_i = (1 - \dot L_i/c)$.  Other $y_{ij}$ are
determined from equations (1) and (2) by cyclic index permutation.

Inserting these into the equations for $\alpha$ and $\rho$ we can
verify that the laser phase/frequency noises cancel and that the
optical bench noises and USO noises are, for reasonable parameters of
SyZyGy, suppressed to acceptable levels.  Noting that $\Delta_2 \Delta_3 \simeq \Delta_1$
and $\omega_{23} - \omega_{13} = \nu_1 \Delta_2 - \nu_2 \Delta_1 \simeq -\omega_{31}$,
and its permutations,
we then obtain
the following responses to proof-mass, USO, and optical path noises in 
X and $\rho$ or $\sigma$.

\begin{eqnarray}
X^{noise} & = & X^{proof \ mass} + X^{USO} + X^{opt. \ path}
\nonumber \\
&  = &  - 2 (\nu_1/\nu_0) (\hat n \cdot \vec v_1 - \hat n \cdot \vec v_{1;11})
+ 2 (\nu_2/\nu_0) (\hat n \cdot \vec v_{2;3} - \hat n \cdot \vec v_{2;322})
\nonumber \\
& & + 2 (\nu_3/\nu_0) (\hat n \cdot \vec v_{3,2} - \hat n \cdot \vec v_{3;233})
\nonumber \\
& & - 2 (\omega_{31}/\nu_0) (\hat n \cdot \vec v_1 - \hat n \cdot \vec v_{1;22})
- 2 (\omega_{21}/\nu_0) (\hat n \cdot \vec v_1 - \hat n \cdot \vec v_{1;33})
\nonumber \\
& & + 2 (\omega_{32}/\nu_0) (\hat n \cdot \vec v_{2;3} - \hat n \cdot \vec v_{2;322})
+ 2 (\omega_{23}/\nu_0) (\hat n \cdot \vec v_{3;2} - \hat n \cdot \vec v_{3;233})
\nonumber \\
& &  + (\omega_{31}/\nu_0) (Q_1 - Q_{1;22}) - (\omega_{21}/\nu_0) (Q_1 - Q_{1;33})
\nonumber \\
& &  + (\omega_{32}/\nu_0) (Q_{2;3} - Q_{2;12})
- (\omega_{23}/\nu_0) (Q_{3;2} - Q_{3;13})
\nonumber \\
& & + y_{32;322}^{opt. \ path} - y_{23;233}^{opt. \ path} + y_{31;22}^{opt. \ path}  - y_{21;33}^{opt. \ path}
\nonumber \\
& & + y_{23;2}^{opt. \ path} - y_{32;3}^{opt. \ path} +y_{21}^{opt. \ path} - y_{31}^{opt. \ path}
\label{eq:5}
\end{eqnarray}

\begin{eqnarray}
\rho^{noise} & = & \rho^{proof \ mass} + \rho^{USO} + \rho^{opt. \ path}
\nonumber \\
&  = & 2 (\omega_{31}/\nu_0) [ \hat n \cdot \vec v_3 - \hat n \cdot \vec v_{1;2}]
\nonumber \\
& &  - (\omega_{31}/\nu_0) (Q_3 - Q_{1;2})
\nonumber \\
& & + y_{31}^{opt. \ path} - y_{23}^{opt. \ path} - \Delta_2 y_{31;2}^{opt. \ path} 
\label{eq:6}
\end{eqnarray}

\noindent
Under the assumptions that all noises are independent, that optical
path noise only depends on the distance between spacecraft pairs, and that
for proof mass noise in X terms with coefficients $\omega_{ij}/\nu_0$
are negligible compared with terms having coefficients of order unity,
the
spectra of the noises in X and $\rho$ are:

\begin{eqnarray}
S_{X}(f) & = & S_{X}^{proof \ mass} + S_{X}^{USO} + S_{X}^{opt. \ path}
\nonumber \\
&  \simeq &  16 [\sin^2(2 \pi f L_1) + \sin^2(2 \pi f L_2) + \sin^2(2 \pi f L_3)] S^{proof \ mass}
\nonumber \\
& &  + [ 4 (\omega_{32}^2 + \omega_{31}^2)/\nu_0^2 \ \sin^2(2 \pi f L_2) + 4 (\omega_{23}^2 + \omega_{21}^2)/\nu_0^2 \
 \sin^2(2 \pi f L_3)
\nonumber \\
&  & -8 (\omega_{21} \omega_{31})/\nu_0^2 \  \cos(2 \pi f (L_2 - L_3)) \  \sin(2 \pi f L_2) \  \sin(2 \pi f L_3)] S^{USO} 
\nonumber \\
& & + 8 \sin^2(2 \pi f L_3) S_{23}^{opt. \ path} + 8 \sin^2(2 \pi f L_2) S_{31}^{opt. \ path}
\label{eq:7}
\end{eqnarray}

\begin{eqnarray}
S_{\rho}(f) & = & S_{\rho}^{proof \ mass} + S_{\rho}^{USO} + S_{\rho}^{opt. \ path}
\nonumber \\
&  = &  8 (\omega_{31}^2/\nu_0^2) S^{proof \ mass} + 2 (\omega_{31}^2/\nu_0^2) S^{USO}
\nonumber \\
& & +  S_{13}^{opt. \ path} +  S_{23}^{opt. \ path} +  S_{31}^{opt. \ path}  
\label{eq:8}
\end{eqnarray}

\section{Gravitational Wave Signal Response}

The gravitational wave signal response in TDI combination X was
given for a general triangle in equation (24) of reference \cite{1}.
The SyZyGy configuration leads to considerable simplification.  Taking
the limit as the general triangle is distorted to a line, using $L_1 =
L_2 + L_3$, and remembering that $\hat n = \hat n_1 = -\hat n_2 = -\hat
n_3$, we obtain a six pulse response with a single amplitude depending
only on the orientation of the linear array with respect to the source
direction:

\begin{eqnarray}
X^{gw}   & = &  2 \ \hat k \cdot \hat n  \ [ \Psi(t - L_2 + L_2 \ \hat k \cdot \hat n) - \Psi(t)
\nonumber \\
& &  + \ \Psi(t - L_3 - L_3 \ \hat k \cdot \hat n) - \Psi(t - L_1 - L_2 - L_3 \ \hat k \cdot \hat n)
\nonumber \\
& &  + \ \Psi(t - L_1 - L_2 - L_3) - \Psi(t - L_1 - L_3 + L_2 \ \hat k \cdot \hat n)]
\label{eq:9}
\end{eqnarray}


\noindent 
where $\hat k$ is the gravitational wavevector and $\Psi$ is (see \cite{6,7})

\begin{eqnarray}
\Psi(t) & = & {1 \over 2} \ \
{{{\hat n} \cdot {\bf h}(t) \cdot {\hat n}} \over  { 1 - ({\hat k} \cdot {\hat n})^2}}
\label{eq:10}
\end{eqnarray}

\noindent
for a transverse traceless gravitational wave  $\bf{h}$(t) = $\left[
h_+(t) \  {\bf{e_+}} + h_{\times}(t) \ {\bf{e_{\times}}} \right]$.  Note that $X^{gw}$ is
zero for waves incident at right angles to SyZyGy ($\hat k \cdot \hat n$ = 0) and
for waves parallel or antiparallel with it ($\hat k \cdot \hat n$ = $+1$ or $-1$).

Although SyZyGy will not operate exclusively in the long-wavelength limit (LWL - $2 \pi f L_1 << 1$),
analytical results at low-frequency are useful.  Expanding the gravitational wave ${\bf h}$ in
series of spatial derivatives and substituting into $X^{gw}$ gives

\begin{eqnarray}
X^{gw} \to - \mu  L_1 L_2 L_3  \  \hat n \cdot {\bf{h^{'''}}} \cdot \hat n
\label{eq:11}
\end{eqnarray}

\noindent 
where $\mu = \hat k \cdot \hat n$.

\section{Gravitational Wave Sensitivity}

Figure 4 shows root-mean-square $X^{gw}$, averaged over
polarization and the celestial sphere \cite{1, 2}, for 3 armlength
cases:  (a) $L_2 = L_3 = 16$ light-seconds, (b) $L_2 = 7, L_3 = 9$
light-seconds, and (c) $L_2 = 0.7, L_3 = 0.9$ light-seconds.  Notice that
in the LWL $X^{gw} \propto f^3$, as expected from the LWL expansion 
above.

Figure 5 shows the noise power spectra of X, using proof mass and
optical path noises only, for the same armlength cases as Figure 4.
(USO can be made negligible compared to the other noises through
trajectory control:  if the maximum relative speed is kept below 0.2
m/sec--either through trajectory design or by periodically using
thrusters to adjust the maximum Doppler shift--USO noise will not
contribute.) In constructing the curves in Figure 5 we used the
same instrumental spectra as in our previous sensitivity analysis of
LISA.  The proof masses
were assumed to have independent noises with one-sided acceleration
spectral density $3 \times 10^{-15}$ m sec$^{-2}$ Hz$^{-1/2}$ (this
corresponds \cite{2,3} to a one-sided spectral density of relative
Doppler $S_y^{proof\ mass} = 2.5 \times 10^{-48} $[f/1 Hz]$^{-2}
$Hz$^{-1}$.  Following previous practice \cite{2,3}, we approximately
account for all optical path noise, including beam pointing noise (see
\cite{2} and reference \cite{8}, Table 4.1), on a given link with
$S_y^{optical\ path} = 1.8 \times 10^{-37}$ [f/1 Hz]$^2 ($r$/ 5 \times 10^9
$m$)^2 $Hz$^{-1}$, where r is the distance between spacecraft on that
link.  The aggregate noise spectrum plotted uses these spectra and the
X transfer function (see equations 4 and 6).  

The GW sensitivity is proportional to the ratio of the rms noise to the
rms signal.  Figure 6 shows the sensitivity of SyZyGy expressed in the
conventional way:  sensitivity to isotropic sinusoidal gravitational
radiation (SNR = 5 in one-year integration),  $5 [S_X(f)\ B]^{1/2}$/rms
$X^{gw}$(f), where B = 1 cycle/year, for the three armlength cases of
Figures 4 and 5.

\section{Discussion}

SyZyGy offers some technical simplifications compared with other
configurations.  Each spacecraft utilizes a single proof mass, and laser
metrology is needed only at the central spacecraft.  Because of the low relative
velocities afforded by the SyZyGy orbit, the complexity needed for
USO noise elimination (laser beam modulation and additional calibration data)
\cite{3} is not required by SyZyGy.
The angles between the arms are much more constant in time than for
LISA, so time-dependent articulation of the telescopes is not
required.  SyZyGy combinations $\rho$ and $\sigma$ have exactly zero GW
response and can thus be used with no ambiguity to assess on-orbit
noise performance.  Deliberate resetting of arm lengths (i.e.,
different arm lengths in different portions of a long mission) is
easier in a linear configuration and could be used to tune SyZyGy's
best response over its mission lifetime.

For similar scale size, SyZyGy achieves both its best sensitivity  and
high-frequency performance comparable to those of LISA; at low-frequencies,
SyZyGy has poorer sensitivity.  This suggests applicability in the
Fourier band between those of LISA and ground-based interferometers.  Little
attention has been given to the $\sim 0.1 - 10$ Hz band.   
The work that has been done suggests it to be relatively free of
foreground (astrophysical) GW sources \cite{9}.  A SyZyGy with $L_1
\sim$ 1 light-second could test this with $5 \sigma$ sensitivity $\simeq
2.5 \times 10^{-23}$ in a year's integration.

\section*{Acknowledgments}
This research was performed at the Jet Propulsion Laboratory,
California Institute of Technology, under contract with the National
Aeronautics and Space Administration.

\clearpage

FIG. 1.  SyZyGy configuration.  Unit vectors $\hat n_i$ point between
spacecraft pairs with the indicated orientations.  The arrows on the 
$L_i$ indicate the sense of the un-primed light times.  Each spacecraft
has one optical bench and one proof mass.

FIG. 2.  Signal routing and readouts on the central spacecraft, 1.

FIG. 3.  Signal routing and readouts on spacecraft 3, one of the end spacecraft.

FIG. 4.  RMS gravitational wave response of SyZyGy, TDI combination
X, as a function of
Fourier frequency for three cases:  $L_2 = L_3 = 16$ light-seconds;
$L_2 = 7,  L_3 = 9$ light seconds; and $L_2 = 0.7,  L_3 = 0.9$
light seconds.

FIG. 5.  One-sided noise spectra of X as a function of Fourier
frequency.  One-sided spectra of proof mass noise and optical path
noise are taken to be the the same as we have used in previous
calculation of LISA sensitivity \cite{2, 3}: respectively, $3 \times
10^{-15}$ m sec$^{-2} $Hz$^{-1/2}$ and $20 \times 10^{-12} \times ($r$/ 5
\times 10^9 $m$)$ m Hz$^{-1/2}$, where r is the distance between
spacecraft.  Center frequencies of the lasers are assumed to be equal
and a maximum Doppler shift (longest arm case) of 0.2 m/sec is assumed.  Solid
lines:  spectra of X including shot and proof mass noises.  Dashed
lines:  contribution to X noise spectrum from an uncalibrated USO 
having one-sided spectrum of fractional frequency
fluctuations $8 \times 10^{-27}/$[f/1Hz] (see text).

FIG. 6.  Gravitational wave sensitivity of X for SyZyGy (SNR = 5 in a
one year integration, sky averaged, sinusoidal signals).  Arm lengths,
gravitational wave responses, and noise spectra are as in Figures 4-5.

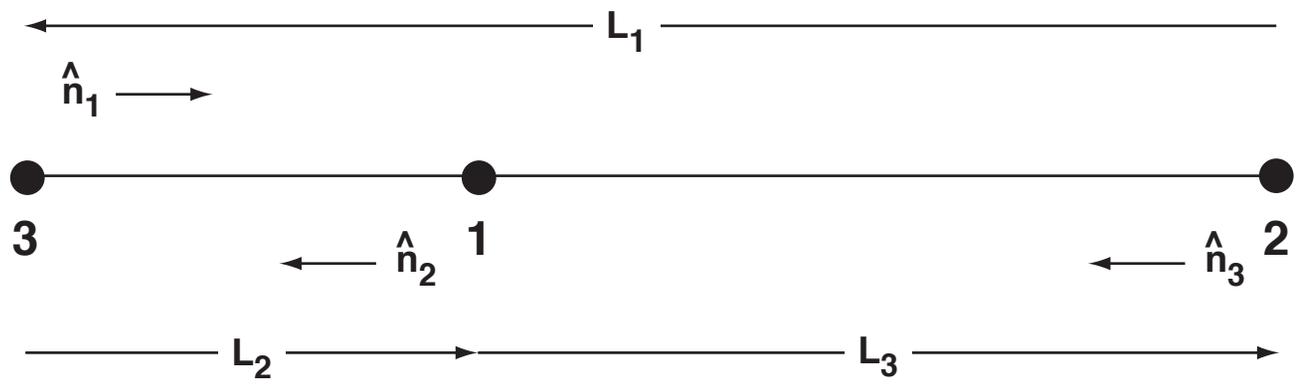
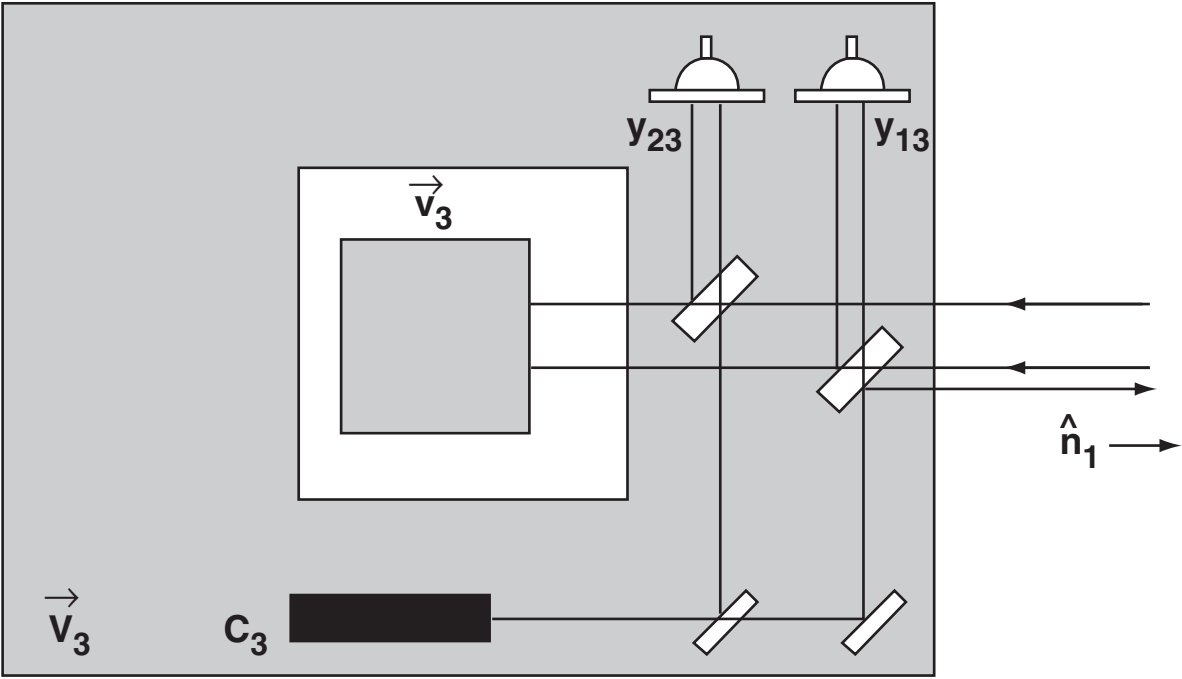
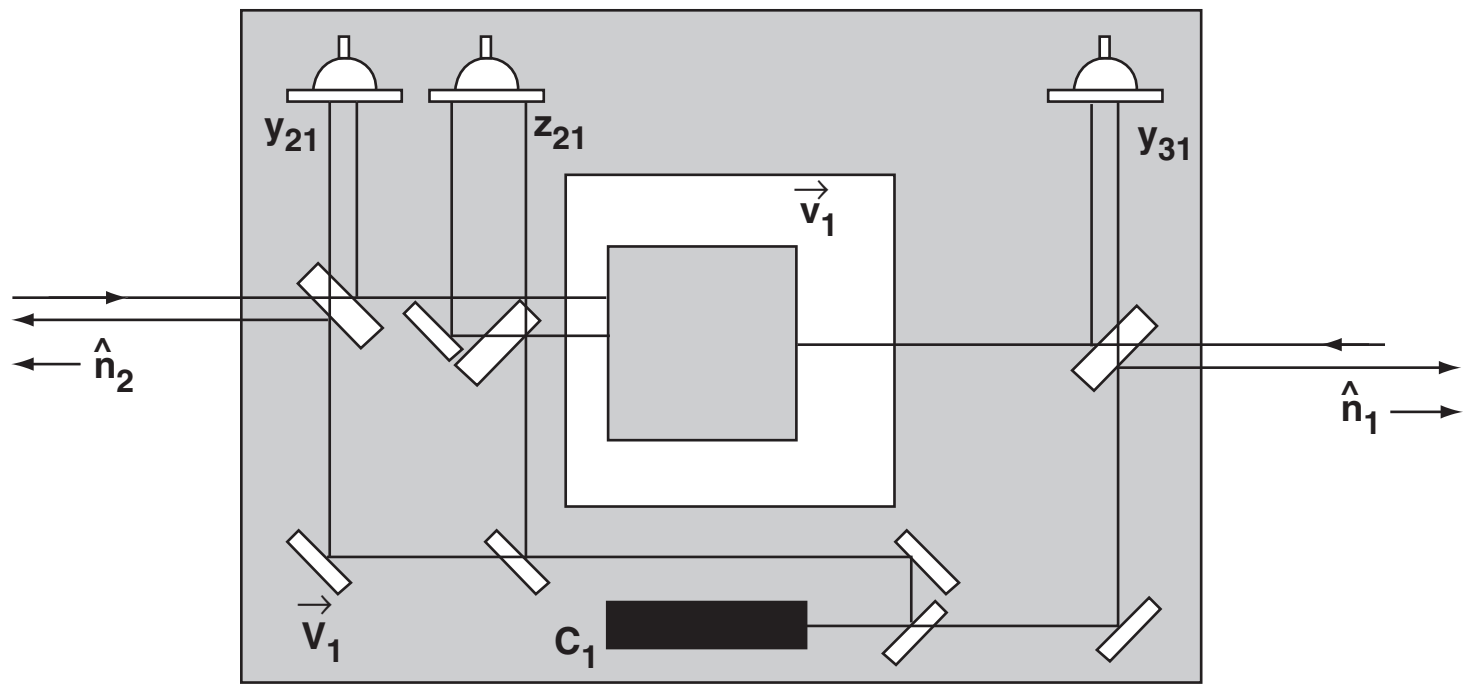
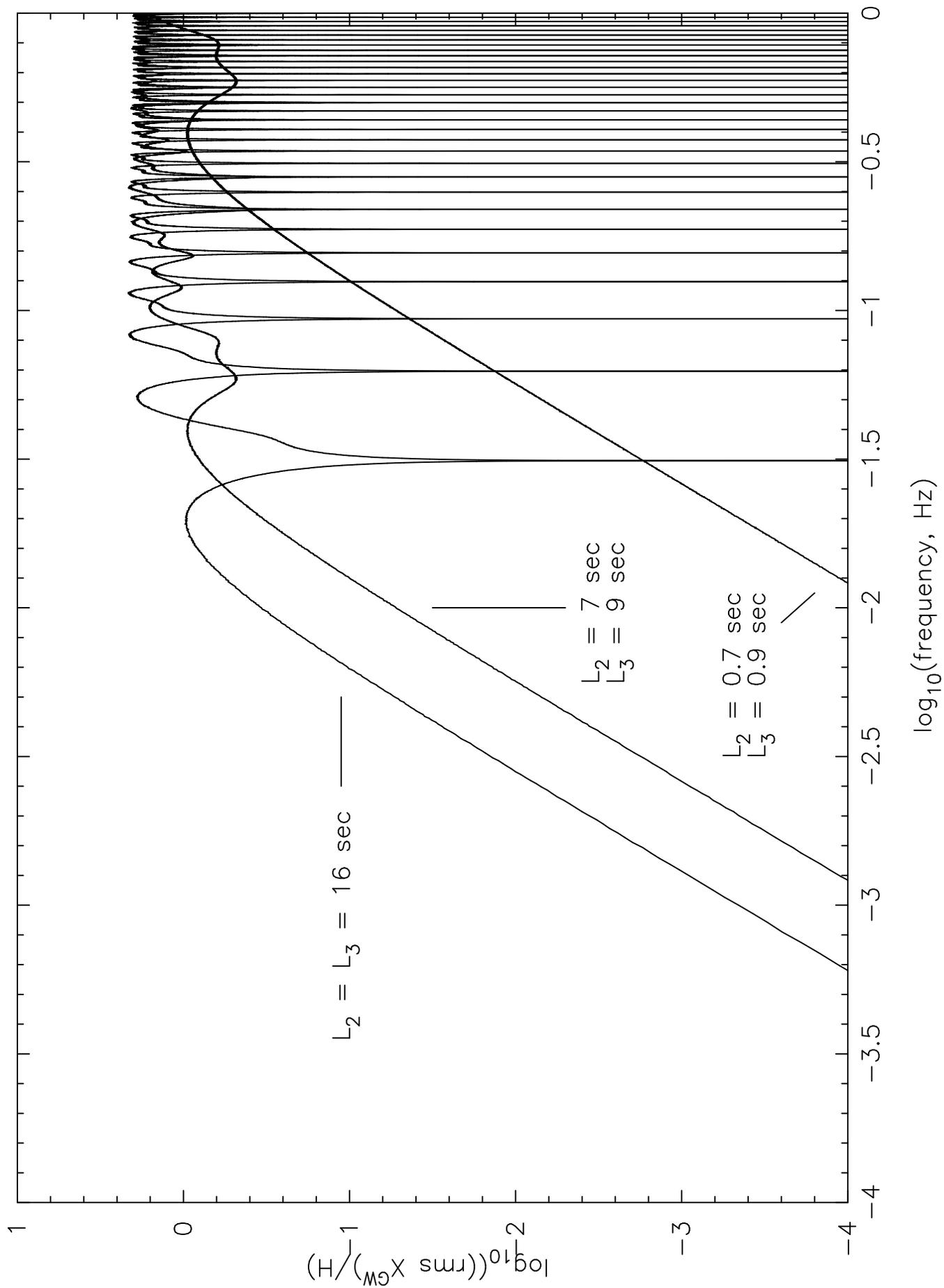
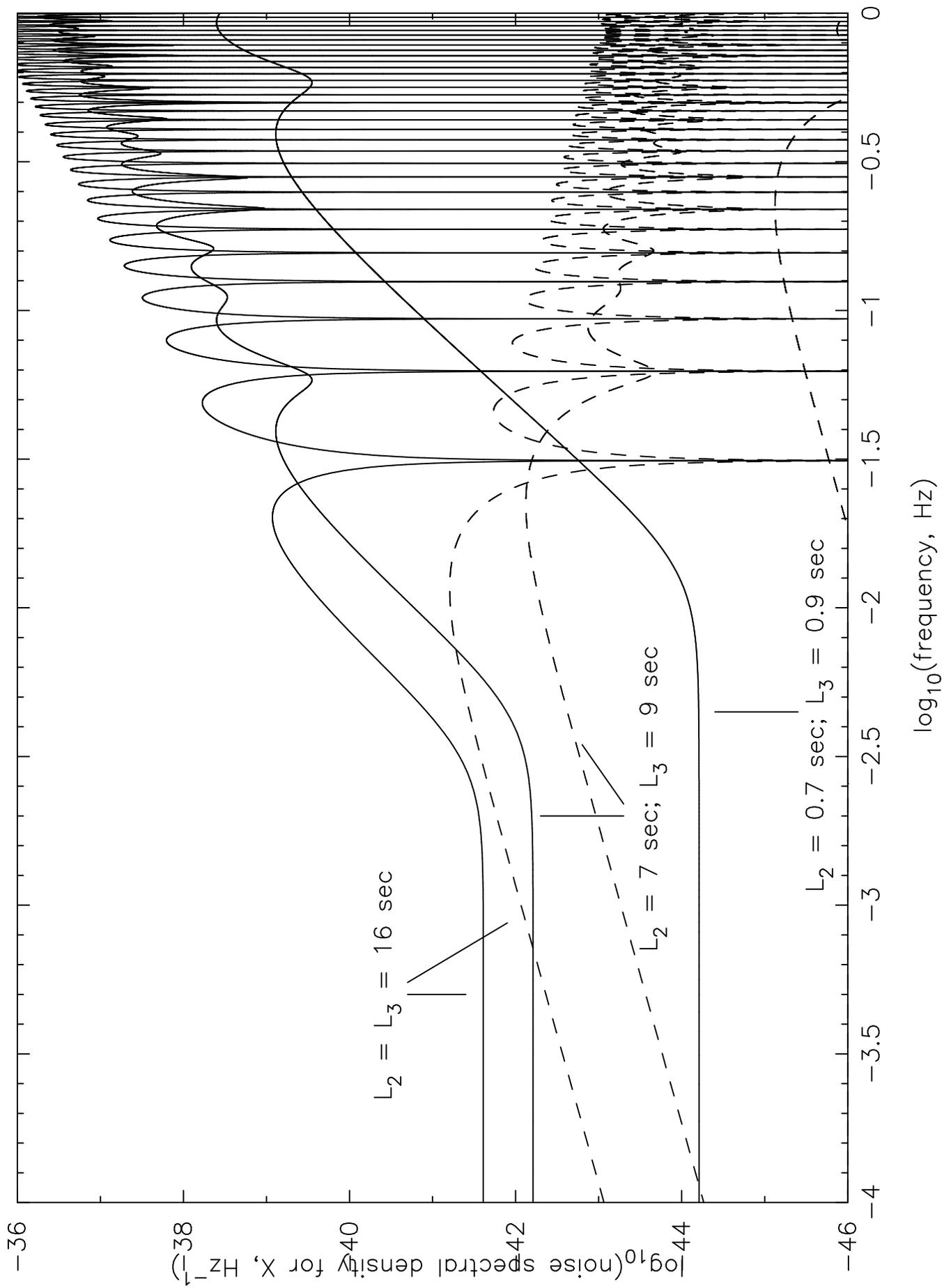
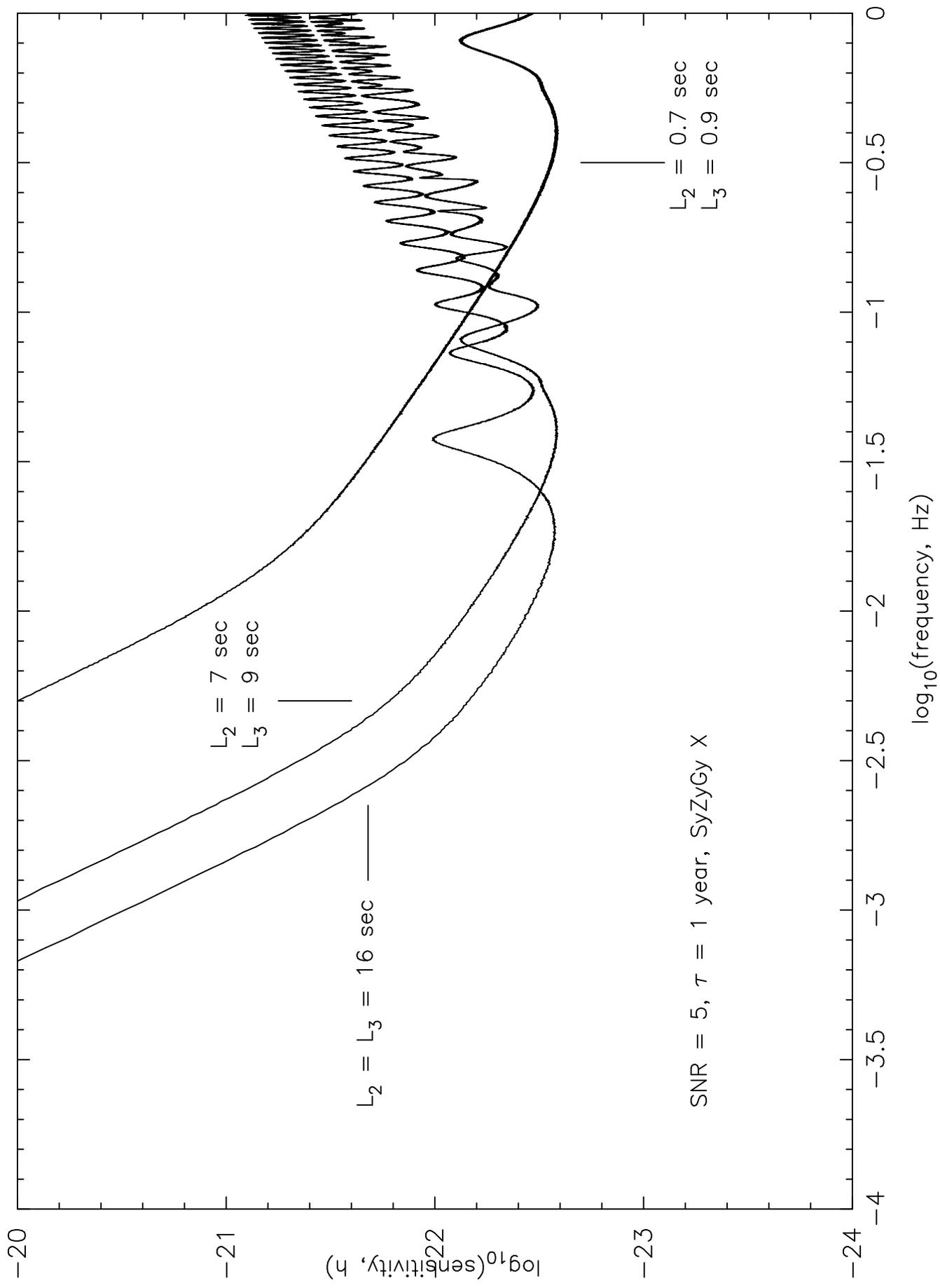
\end{document}